# Emergence of ferroelectric topological insulator as verified by quantum Hall effect of surface states in (Sn,Pb,In)Te films

Ryutaro Yoshimi[1, 6*], Ryosuke Kurihara[2, 7], Yoshihiro Okamura[3], Hikaru Handa[3], Naoki Ogawa[1], Minoru Kawamura[1], Atsushi Tsukazaki[4], Kei S. Takahashi[1], Masashi Kawasaki[1, 3], Youtarou Takahashi[3], Masashi Tokunaga[2], Yoshinori Tokura[1, 3, 5]

[1] RIKEN Center for Emergent Matter Science (CEMS), Wako 351-0198, Japan.

[2] The Institute for Solid State Physics, The University of Tokyo, Kashiwa, Chiba, 277-8581, Japan.

[3] Department of Applied Physics and Quantum-Phase Electronics Center (QPEC), The University of Tokyo, Tokyo 113-8656, Japan

[4] Institute for Materials Research, Tohoku University, Sendai 980-8577, Japan

[5] Tokyo College and Department of Applied Physics, The University of Tokyo, Tokyo 113-8656, Japan

[6] Present address: Department of Advanced Materials Science, The University of Tokyo, Kashiwa 227-8561 Japan

[7] Present address: Department of Physics and Astronomy, Tokyo University of Science, Noda, Chiba 278-8510, Japan

* Corresponding author: r-yoshimi@edu.k.u-tokyo.ac.jp

**ABSTRACT**

Emergent phenomena arising from nontrivial band structures based on topology and symmetry have been attracting keen interest in contemporary condensed-matter physics. Materials such as SnTe and PbTe are one such example, which demonstrate a topological phase transition while showing ferroelectric instability derived from their rock-salt structure. The ferroelectricity can lift the valley degeneracy, enabling the emergence of the $Z_2$ topological insulator phase, although its observation in transport phenomena remains elusive. Here, we report magnetotransport properties of ferroelectric (Sn,Pb)Te thin films with finely-controlled Fermi levels via In doping. We identified the ferroelectric topological insulator phase from the observations of the quantum Hall states with filling factors of $v = 1$, 2 and 3 with both spin- and valley-degeneracy lifting. The electronic states are two-dimensional, indicating the ferroelectricity-induced topological surface states with a single Dirac cone. The finding of the new topological state with ferroelectricity will further expand the field of topological physics and advance the development of functional properties, such as topological nonlinear photonics and nonreciprocal transport with memory effect.

**MAIN TEXT**

Exploration of novel physical phenomena originating from nontrivial band topology has been garnering great attention in the field of condensed matter science and technology [1–5]. The topological phase transition between topological insulator, Weyl/Dirac semimetal and normal insulator as driven by band inversion and symmetry breaking are described via the universal phase diagram predicted by Murakami and coworkers [6,7]. In that framework, topological semimetals such as Weyl semimetal or nodal line semimetals are expected to emerge in the systems with broken inversion symmetry between normal insulator (NI) and topological insulator (TI) or topological crystalline insulator (TCI) phases. The time-honored semiconductors PbTe-SnTe system has gained revived interest as a promising platform for exploring nontrivial band topology and symmetry breaking based on the Murakami's diagram [8–12]. SnTe is a semiconductor with inverted band structure, exhibited by the TCI that possesses gapless surface states protected by mirror symmetry of the crystal structure [8,9], whereas PbTe is a normal insulator (NI) known for photoconductive and thermoelectric properties [13–15]. In addition, these materials exhibit inherent ferroelectric instability originating from the rock-salt structure [16–22]. The ferroelectric instability facilitates the emergence of Weyl semimetal phase near the band inversion (Fig. 1(a)), displaying exotic transport behaviors such as non-linear Hall effect associated with Berry curvature dipole and enhancement of anomalous Hall conductivity [23–26].

The ferroelectric polarization lifts the valley degeneracy via the breaking of spatial inversion symmetry, which may also affect the topological phase of this materials class. In the TCI phase, four gapless surface states appear at one $\bar{\Gamma}$ and three $\bar{M}$ points on the (111) surface (Fig. 1(b)). These surface states are derived from the bulk band gap located at the L points in the first Brillouin zone, which are crucial in determining the topological phase. Without ferroelectric polarization, the band inversion should occur equivalently at all the L points, resulting in the topological phase transition as discussed above. In contrast, when ferroelectric order is established with the electric polarization along the [111], the degeneracy at the four L points is lifted into one and three (Fig. 1(d)). In this situation, while a TCI appears with an even number of surface states as protected by crystal mirror symmetry, a $Z_2$ TI state protected by time-reversal symmetry can appear with an odd number of gapless surface states [27] (Fig. 1(c)). In fact, emergence of $Z_2$ TI is theoretically predicted for SnTe by the introduction of lattice strain [27,28], and is experimentally confirmed in a similar materials system (Pb,Sn)Se doped with Bi by angle-resolved photoemission spectroscopy [29,30]. The emergence of these novel topological phases, particularly when the Fermi level ($E_F$) is tuned close to the Dirac point of the surface state, could facilitate the observation of transport and photonic properties highly sensitive to symmetry and topology. In particular, the thin film samples, which are susceptible to lattice strain to control the electrical polarization vector, are suited for exploring the above physics and functions.

Here, we investigate magnetotransport properties of (Sn,Pb)Te thin films doped with In (indium). We focused on a specific composition range that shows characteristic features of ferroelectricity in optical responses and high-mobility charge transport with low carrier density. By utilizing pulsed high field measurements up to 55 T, we have revealed that the (Sn,Pb)Te thin films exhibit the quantum Hall effect of the Landau level filling factor $v$ = 1, 2 and 3 with almost zero residual resistivity at $v$ = 1 and 3. In addition, the Shubnikov de-Haas oscillation (SdH) shows the clear two-dimensional angular dependence. From these observations, we conclude that the electronic states that host the quantum Hall effect are two-dimensional surface states of topological insulator with full spin- and valley-polarization as induced by the ferroelectricity.

We grew 40 nm-thick (111)-oriented $(Sn_xPb_{1-x})_{1-y}In_yTe$ (SPIT) thin films on InP (111) substrates (Fig. 1(e)) by molecular beam epitaxy (MBE). The chemical compositions of the samples in this study are controlled with doping of Sn ($x$ = 0.16 - 0.19) and In ($y$ = 0 - 0.17) into a parent compound rock-salt PbTe.

The lattice constant of PbTe (6.46 Å at 298 K) [31] is approximately 10% larger than that of InP (5.86 Å) [32], so that we inserted 2-nm thick SnTe buffer layer (lattice constant 6.31 Å) [31] beneath the SPIT layer and could obtain smooth (111)-oriented SPIT thin films (see Methods). The previous experiment shows that the band gap nearly closes around this composition range [12]. In addition, optical second-harmonic-generation (SHG) measurement for the SPIT thin film with $(x, y) = (0.16, 0.065)$ verifies the presence of electric polarization along out-of-plane direction [111] (Fig. 1(f)), and THz spectroscopy reveals the softening of transverse optical phonons with the decreasing of temperature, being reminiscent of the inherent ferroelectricity (Fig. 1(g)). Figure 1(h) displays the temperature dependence of $\rho_{xx}$ for the identical thin film. It exhibits a semiconducting behavior above 200 K, while it changes to a metallic behavior below 200 K, accompanied by a significant reduction in resistivity. This behavior is similar to the temperature dependence of resistivity in other low-carrier TI compounds $Bi_2Te_2Se$ [33,34]; at high temperatures, carriers in three-dimensional bulk states contributes to the semiconducting behavior, while metallic conduction on the surface becomes dominant at low temperatures. This is also consistent with the increase in mobility as the temperature decreases (Fig. 1(h)). A closer examination below 30 K (Fig. 1(i)) further reveals a sharp drop in resistivity around 20 K, accompanied by a doubled increase in carrier density. Although no anomalies are seen in SHG signals (Fig. 1(f)) nor in optical phonon resonance frequencies (Fig. 1(g)) at around 20 K, we speculate that the low-temperature variation in resistivity and carrier density is relevant to a further change in the electronic structure around the Fermi level ($E_F$), as argued later in terms of the subtle charge transfer between the gapped bulk bands and the surface Dirac band. Despite the doubled increase in the carrier density, the carrier density of $1.5 \times 10^{17}$ cm$^{-3}$ is still low enough to examine the quantum transport near the band crossing point. Thus, we focus on the low-resistivity region that appears below 15 K in this experiment and examine the electronic structure by means of magnetotransport properties.

Figure 2(a) presents the magnetic field dependence of longitudinal and Hall resistance, $R_{xx}$ and $R_{yx}$, for the SPIT thin film with $(x, y) = (0.16, 0.065)$ measured by pulsed magnetic fields up to $B = 55$ T at $T = 1.4$ K. As the magnetic field increases, $R_{xx}$ exhibits SdH oscillations, while $R_{yx}$ shows plateaus at 16 T, 28 T, and above 40 T. The values of the $R_{yx}$ at these plateaus are respectively denoted by $h/3e^2$, $h/2e^2$, and $h/e^2$, indicating the emergence of integer quantum Hall (QH) effect where $|R_{yx}|$ is quantized in a unit of $h/\nu e^2$. Here, $h$, $e$ and $\nu$ represent the Planck constant, elementary charge and the Landau level filling factor, respectively. Corresponding to the formation of the QH plateaus, dips appear in $R_{xx}$. While $R_{xx}$ drops to zero at the odd-integer QH states ($\nu = 1$ and 3), a residual resistivity is observed at even states (*e.g.*, $\nu = 2$). This trend becomes clearer when plotting $R_{xx}$ against $1/B$, the period of the SdH oscillations. As shown in Fig. 2(b), in addition to the $\nu = 1, 2,$ and 3 states, odd-integer QH states with $\nu = 5$ and 7 are clearly resolved, whereas even-integer states with $\nu \geq 4$ cannot be discerned. The reason why the even-integer states are rather unstable compared with the odd-integer states can be elucidated by the top and bottom surface degrees of freedom of the TI surface states, as argued below.

The zero longitudinal resistance at $\nu = 1$ and 3 indicates that there is no density of states at the Fermi energy. In addition, the observed QH states with $\nu = 1, 2$ and 3 mean the lifting of both valley and spin degeneracies or the fully spin- and valley-polarized states. From these points, we propose that the observed QH effect is ascribed to the emergence of a TI phase driven by ferroelectricity and the associated surface states featuring a single-valley and spin-polarized Dirac dispersion. Due to the ferroelectricity along the [111] direction that is perpendicular to the film plane, the degeneracy of four bulk L points is lifted into one that is along [111] and the other three; hereafter we refer to $L_1$ and $L_2$ - $L_4$, respectively. When the band inversion takes place at the $L_1$ but does not at $L_2$ - $L_4$, only one gapless surface state appears at $\bar{\Gamma}$ point in the surface Brillouin zone, *i.e.*, the projection of $L_1$ on the (111) surface, and the other three states at $\bar{M}$ remain gapful (Fig. 2(c)). Moreover, the ferroelectric polarization perpendicular to the plane causes the energetic difference between top and bottom surfaces (Fig. 2(d)). In that case, the QH states appear at $v = (m + 1/2) + (m' + 1/2) = m + m' + 1$ with $m$ being integer as the sum of the top and bottom surfaces, meaning that $v$ can take every integer value. This is because a single Dirac cone exhibits the half-integer QH states, *i.e.* $v = m + 1/2$, due to the Berry phase, as is well known for the QH effect in graphene [35,36] (a schematic energy diagram shown in Fig. 2(e)). Indeed, the emergence of QH states by the energy difference between top and bottom surface states have been demonstrated in a $(Bi,Sb)_2Te_3$ TI thin film as exemplified by the emergence of $v = 0$ QH state [37,38]. In this mechanism, the energy gaps for the even-integer QH states originate from energy difference between top and bottom surfaces regardless of the Landau level indexes. This also supports the present observation of the less pronounced even-integer QH states than the odd-integer states, such as nonzero residual resistivity at $v = 2$ and indiscernibility of $v = 4$ and 6 states; namely in the degenerate limit of the top and bottom surface states, the only $v = 2m + 1$ states would show up. The observed behavior of the QH effect with the independent contributions from the top and bottom surface states excludes the scenario of quantum transport of Weyl semimetal state (see Figs. 1(a) and 1(b)) where the top and bottom surface sates would be connected each other via the Weyl orbit [39,40].

To demonstrate the contribution of two-dimensional surface states more clearly, we measured the angular dependence of the magnetoresistance. As illustrated in the inset of Fig. 3(a) we applied the magnetic field tilted by the angle of $\theta$ from normal to the sample plane. Figure 3(a) shows the magnetic field dependence of SdH oscillations measured for various $\theta$ values. Here, a second derivative with respect to the magnetic field is taken to make the oscillatory component clear. The SdH oscillations diminishes as the magnetic field tilts towards the plane, disappearing above 70 degrees. Performing a fast Fourier transform to determine the oscillation periods $B_{freq}$, which are displayed as lower triangles in Fig. 3(b). By plotting $B_{freq}$ against $\theta$, an excellent fit to $1/\cos\theta$ dependence is revealed as shown in Fig. 3(c). This strongly indicates that the electronic states which host the SdH and QH effect are two-dimensional states at surface or interface.

Next, we argue the role and impact of In doping to tune the $E_F$ in the $(Sn_xPb_{1-x})_{1-y}In_yTe$ (SPIT) thin films on the basis of the proposed electronic structure shown in Fig. 2(c). In Figs. 4(a) and 4(b) the carrier mobility ($\mu$) and density ($n$ for electron-type and $p$ for hole-type carriers) at 2 K are plotted as a function of In concentration ($y$) for the fixed Sn/Pb ratio ($x = 0.16$ for most of the SPIT samples and $x = 0.19$ for one sample). In this Pb-rich (low-$x$) region, the SPIT in the non- or low-In doping ($y$) region, which is referred as region (i), shows the $p$-type behavior with low $\mu$, whereas the SPIT with increase in $y$ is turned to $n$-type ($y > 0.04$) and shows high $\mu$ (4000 - 7000 cm$^2$V$^{-1}$s$^{-1}$) up to $y \sim 0.08$ (region (ii)). With further increase in In content ($y > 0.1$, region (iii)), the $\mu$ decreases to a moderate value (600 - 2000 cm$^2$V$^{-1}$s$^{-1}$). The high-mobility SPIT thin films belonging to the region (ii) show the well-defined SdH oscillation and the QHE as exemplified in Figs. 4(e)-4(g); the magnetic field dependence of $\sigma_{xx}$ and $\sigma_{xy}$ for the SPIT samples with varying In doping ($y$), (e) 0.047, (f) 0.065 and (g) 0.074 at Pb/Sn ratio ($x$) of 0.16, 0.16, and 0.19, respectively. The SPIT sample shown in Fig. 4(f) is the identical to as discussed in the Figs. 1-3. In all the SPIT samples shown here the quantization of the $\sigma_{xy}$ with $v = 1$ and 2 accompanied by zero or dips in $\sigma_{xx}$ is observed, suggesting the QH states of the same physical origin. Some diversity of the QH effect feature for the respective SPIT samples (e) - (g), such as the deteriorated $v = 3$ feature and the almost zero $\sigma_{xx}$ even at $v = 2$, come from the lower carrier density in the sample (e), and from larger energy difference between top and bottom states in (e) as compared with the sample (f) (see also Fig. 2(e)). Note here that the carrier density estimated from the SdH ($n_{SdH}$) with assumption of the fully spin- and valley-polarized 2D Fermi surface shows a good coincidence with the carrier density ($n_{Hall}$) derived by the normal Hall effect based on the single carrier model, as shown in Fig. 4(d), ensuring the presence of the single Dirac cone at the $\bar{\Gamma}$ point of the surface Brillouin zone of the respective top and bottom surfaces in this ferroelectric TI.

All the above features lead us to speculate the critical compositional-tuning mechanism of $E_F$. With In doping, the relative position of $E_F$ appears to shift higher in the surface electronic structure composed of the one gapless Dirac dispersion with spin-momentum locking at the $\bar{\Gamma}$ point and the three gapped bands at the three $\bar{M}$ points, as schematically shown in Fig. 4(c). In the *p*-type region (i), $E_F$ is likely to position not only at the Dirac dispersion but also at the trivial valence bands around the valleys $\bar{M}_{1-3}$ which may lead to the low mobility as observed. In the *n*-type region (ii), $E_F$ crosses only the Dirac dispersion, producing the high mobility of *n*-type carriers and hence the QH effect characteristic of the spin- and valley-polarized Dirac state. We deduce the $E_F$ position measured from the Dirac point ($E_F$ - $E_D$) using $n_{Hall}$ and Fermi velocity obtained from THz magneto-optical Faraday rotation measurement. As shown in Fig. 4(b), $E_F$ - $E_D$ is approximately 30-80 meV, which is consistent with spectroscopic results. In the *n*-type region (iii), the moderately reduced mobility is perhaps due to the too-high $E_F$ level on the Dirac cone or due to the co-crossing of the trivial conduction bands around the valleys $\bar{M}_{1-3}$. Thus, the tuning of $E_F$ is critical to host the charge quantum transport properties originating from the genuinely topological state realized at $E_F$. Because of the lifting the valley degree of freedom due to the ferroelectric polarization and the compositional $E_F$-tunability by In doping, the finding of such a sweet spot region becomes possible in the present SPIT system.

In summary, we have examined the quantum transport in $(Sn_xPb_{1-x})_{1-y}In_y$Te thin films whose composition was controlled to host the high carrier mobility and the ferroelectricity. Under magnetic fields applied in the ferroelectric-polarization direction, the quantum Hall (QH) effect appears at the Landau level filling factor $v = 1$, 2 and 3. From the observed spin- and valley-polarized features of QH states as well as the angular dependence of Shubnikov de-Haas oscillation, we conclude that the topological insulator emerges in this system, where the valley-selective band inversion is likely to occur due to the ferroelectric polarization along the [111] axis, *i.e.* normal to the film plane. The systematic change in the transport including QH effect is observed in the samples with varying In composition as a tuning knob of the Fermi level, demonstrating the presence of the topological insulator phase around some composition range where only the one valley originating from the bulk L points is subject to the band inversion to form the surface gapless Dirac state. The present successful synthesis of the ferroelectric topological insulator with tunable $E_F$ as exemplified by the QH state may pave a way to emergent functional phenomena in a large areal-size thin films, such as topological photonic and spintronic functions.

**FIGURES**

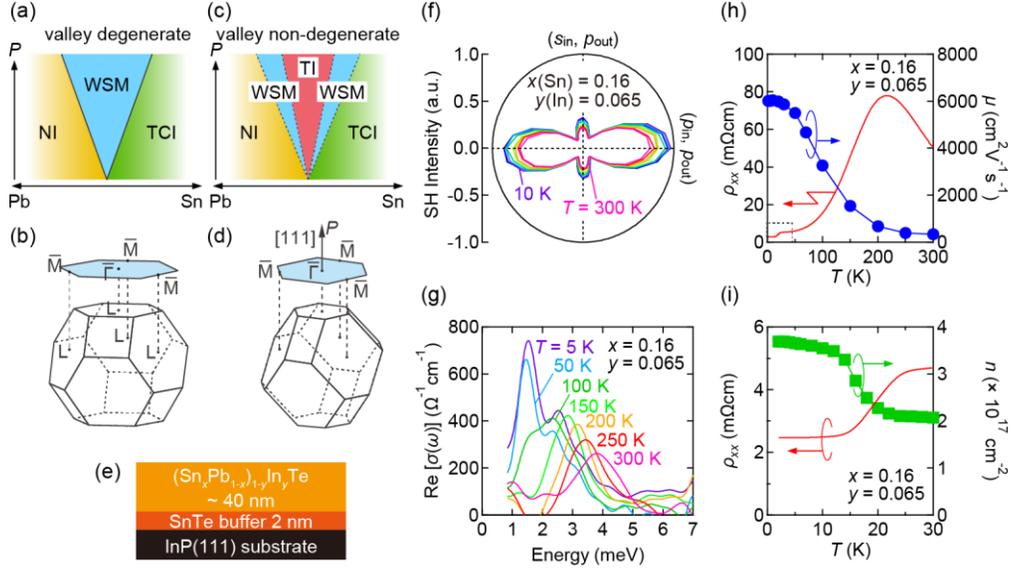

Figure 1 (a, c) Topological phase diagram based on Murakami's scheme (a) without and (c) with [111] polar distortion. (b, d) Sketches of the first Brillouin zone and (111) oriented surface Brillouin zone (b) without and (d) with polar distortion. (e) A sketch of $(Sn_xPb_{1-x})_{1-y}In_yTe$ (SPIT) thin film structure. (f) Incident light polarization dependence of $p$-polarized second harmonic intensity for a SPIT thin film with $(x, y) = (0.16, 0.065)$. (g) Real part of optical conductivity $\sigma(\omega)$ at various temperatures. (h) Temperature dependences of longitudinal resistivity $\rho_{xx}$ and carrier mobility $\mu$. (i) Temperature dependences of $\rho_{xx}$ and electron density $n$ below temperature $T = 30$ K. Temperature dependence of $\rho_{xx}$ is a magnified view of the area enclosed by a dashed box in (h).

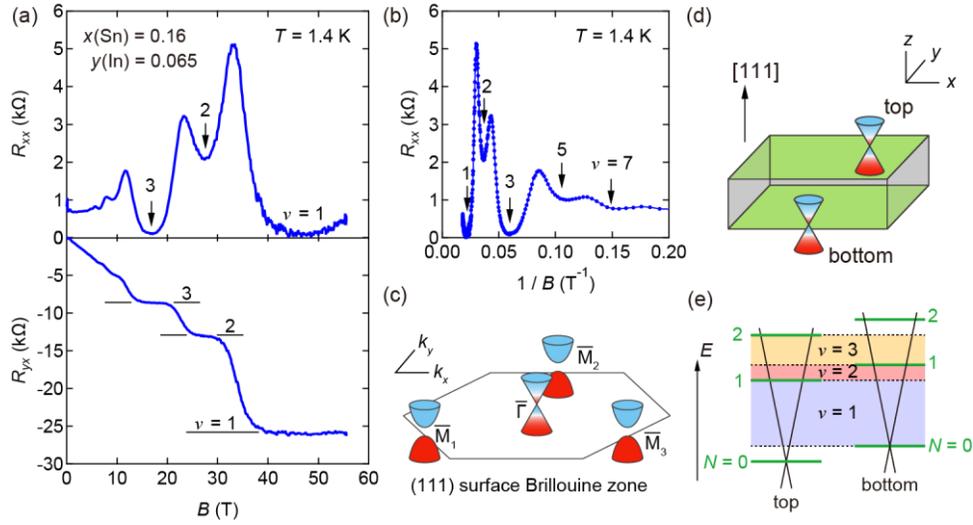

Figure 2 (a) Magnetic field dependence of longitudinal and Hall resistance $R_{xx}$ and $R_{yx}$ for a $(Sn_xPb_{1-x})_{1-y}In_yTe$ thin film with $(x, y) = (0.16, 0.065)$, respectively at temperature $T = 1.4$ K. (b) The magnetic field dependence of $R_{xx}$ plotted against $1/B$. (c) A schematic illustration of surface states at the (111) oriented surface Brillouin zone in the topological insulator phase. (d) A schematic view of surface Dirac states of topological insulator that locates on top and bottom of the thin film. (e) An energy diagram of Landau levels $N$ formed at the top/bottom surface Dirac states, showing the occurrence of quantum Hall states with the Landau level filling factor $\nu = 1$-3.

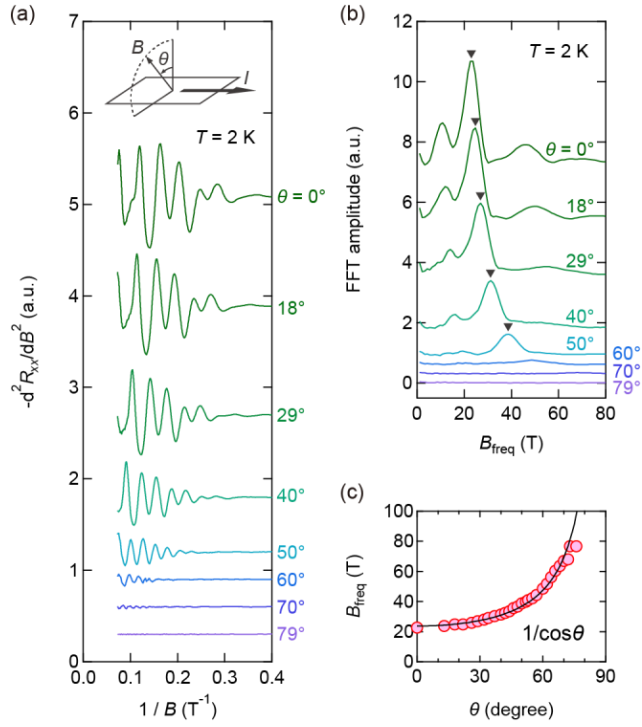

Figure 3 (a) Second derivative of $R_{xx}$ with respect to magnetic field $B$ as a function of $1/B$ at various tilted angle $\theta$. The inset describes an experimental setup. (b) The FFT amplitude of the resistivity oscillation observed in (a). The frequencies of Shubnikov-de Haas oscillations are represented as lower triangles. The offset is shifted for clarity. (c) The resistivity oscillation frequency $B_{freq}$ as a function of tilted angle $\theta$. The solid line represents the fitting line by $1/\cos\theta$.

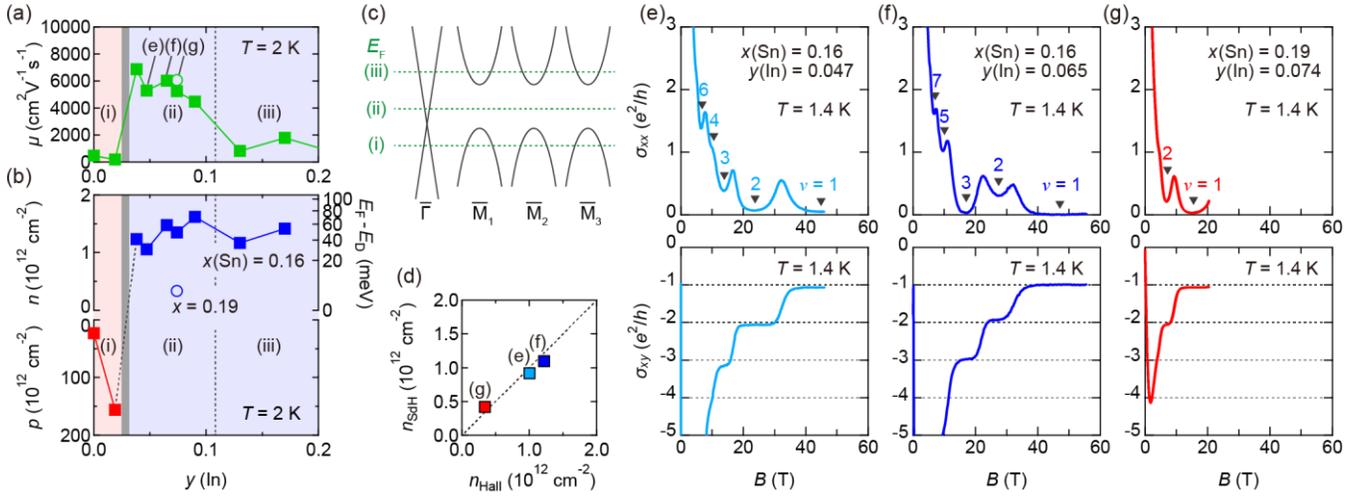

Figure 4 (a, b) Indium doping concentration ($y$) dependence of (a) charge mobility $\mu$, (b) electron and hole density $n$ and $p$, respectively, for $(Sn_xPb_{1-x})_{1-y}In_yTe$ thin film with $x = 0.16$ and $0.19$. (e), (f) and (g) in (a) denotes the sample with $(x, y) = (0.16, 0.047)$, $(0.16, 0.065)$ and $(0.19, 0.047)$ which are displayed in (e), (f) and (g), respectively. One data point for $(x, y) = (0.19, 0.074)$ is shown as an open circle. The right axis in (b) represents the $E_F$ position measured from the Dirac point ($E_F - E_D$). (c) A schematic illustration of the band diagram at the $\bar{\Gamma}$ and $\bar{M}_{1-3}$ points in the surface Brillouin zone (see Fig. 2c) with the typical $E_F$ position corresponding to to the regions of (i) - (iii) (shown in (a) and (b)). (d) Relation between electron density estimated from Shubnikov de Haas oscillation $n_{SdH}$ and that from magnetic field dependence of Hall resistivity $n_{Hall}$. Broken line represents the single-band model relation of $n_{SdH} = n_{Hall}$. (e-f) Magnetic field dependence of two-dimensional longitudinal and Hall conductivity $\sigma_{xx}$ and $\sigma_{xy}$ for $(Sn_xPb_{1-x})_{1-y}In_yTe$ samples with $(x, y) =$ (e) $(0.16, 0.047)$, (f) $(0.16, 0.065)$ and (g) $(0.19, 0.047)$, each representing the quantum Hall effect with the Landau level filling factor $\nu$.

## ACKNOWLEDGEMENTS

We thank Tian Liang, Masataka Mogi, Takuya Nomoto and Ryotaro Arita for fruitful discussion. This research was supported by JSPS KAKENHI Grants (No. 23H01861 and No. 23H05431), and JST CREST (No. JPMJCR1874).